\documentclass[12pt]{article}
\usepackage{epsfig}
{\def\mtiny{\vrule width 0pt}
\def\mrm#1{\mathrm{#1}}
\def\DZ{\relax\ifmmode{D^0}\else{$\mrm{D}^{\mrm{0}}$}\fi}
\def\DONE{\relax\ifmmode{D_1}\else{$\mrm{D}_{\mrm{1}}$}\fi}
\def\DTWO{\relax\ifmmode{D_2}\else{$\mrm{D}_{\mrm{2}}$}\fi}
\def\KZ{\relax\ifmmode{K^0}\else{$\mrm{K}^{\mrm{0}}$}\fi}
\def\BZ{\relax\ifmmode{B^0_d}\else{$\mrm{B}^{\mrm{0}_d}$}\fi}
\def\BZS{\relax\ifmmode{B^0_s}\else{$\mrm{B}^{\mrm{0}_s}$}\fi}
\def\DZS{\relax\ifmmode{D^{*+}}\else{$\mrm{D}^{\mrm{*+}}$}\fi}
\def\DZB{\relax\ifmmode{\overline{D}\mtiny^0}
        \else{$\overline{\mrm{D}}\mtiny^{\mrm{0}}$}\fi}
\def\KZB{\relax\ifmmode{\overline{K}\mtiny^0}
        \else{$\overline{\mrm{K}}\mtiny^{\mrm{0}}$}\fi}
\def\BZB{\relax\ifmmode{\overline{B}\mtiny^0_d}
        \else{$\overline{\mrm{B}}\mtiny^{\mrm{0}_d}$}\fi}
\def\BZBS{\relax\ifmmode{\overline{B}\mtiny^0_s}
        \else{$\overline{\mrm{B}}\mtiny^{\mrm{0}_s}$}\fi}
\def\DZC{\relax\ifmmode{\overline{D}\mtiny^0}
        \else{$\overline{\mrm{D}}\mtiny^{\mrm{0}}$}\fi}
\begin{document}
\title{
\begin{flushright}
\begin{minipage}{1.5in}
	\normalsize
	UCSB HEP 99-08
\end{minipage}
\end{flushright}
Compilation of $\DZ\!\to\!\DZB$ Mixing Predictions}
\author{Harry Nelson}
\date{\today}
\maketitle
\thispagestyle{empty}
\begin{abstract}
We present a compilation of predictions for
the amplitudes of $\DZ\!-\!\DZB$ mixing.
\end{abstract}

We are not aware of any exhaustive compilation of predictions
for the amplitudes for $\DZ\!\to\!\DZB$.  We have therefore found
it helpful to compile, here,
those predictions that we have 
found in a rudimentary search of the literature.

Our search has depended upon the \texttt{SLAC Spires} data
base, available at the \texttt{URL}:
\begin{center}
\texttt{http://www-spires.slac.stanford.edu/find/hep}.
\end{center}

We keep an annotated table of all of the references that we
have found, hyperlinked where possible into both the \texttt{SLAC Spires}
and \texttt{LANL xxx} data bases, at the URL:
\begin{center}
\texttt{http://hep.ucsb.edu/people/hnn/wrongd/predictions/pred.html}.
\end{center}

For all predictions, it is useful to know how to convert
between the two common standards for quoting the mixing
amplitudes: $x$ and $y$, or $\Delta M$ and $\Delta\Gamma$.

Denote the mean life of the $\DZ$ as $\tau_{\DZ}=415\pm4\,$fs\cite{RPP98}, 
and the full width of the $\DZ$ as $\Gamma_{\DZ}=1/\tau_{\DZ}$.
In the limit of $CP$ conservation, $\Delta M=2M_{12}$ and
$\Delta\Gamma=\Gamma_{12}$, where $M_{12}$ and $\Gamma_{12}/2$ are
the mixing amplitudes through virtual and real intermediate states,
respectively.

Then,
\begin{equation}
\begin{array}{rcl}
x&=&\displaystyle{\Delta M\over\Gamma_{\DZ}}\vspace{1.0mm}\\
 &=&\Delta M\tau_{\DZ}\vspace{1.0mm}\\
 &=&\left[\displaystyle{c\tau_{\DZ}\over\hbar c}\right]\times\Delta M
\vspace{1.0mm}\\
 &=&\left[\displaystyle{
2.998\!\times\!10^{23}\hbox{fm/s}\times415\!\times\!10^{-15}\hbox{s}\over
0.1973\,\hbox{GeV-fm}}\right]\times\Delta M\;\hbox{(GeV)}\vspace{1.0mm}\\
 &=&[6.31\!\times\!10^{11}]\times \Delta M\;\hbox{(GeV)}\vspace{1.0mm}\\
y&=&\displaystyle{\Delta \Gamma\over2\Gamma_{\DZ}}\vspace{1.0mm}\\
 &=&[3.15\!\times\!10^{11}]\times \Delta \Gamma\;\hbox{(GeV)}
\end{array}
\end{equation}

For Standard Model predictions, 
we have not made an effort to update author's limits for new
values of the parameters that describe quark mixing (the CKM matrix
elements), or the mass of the top, or any other, quark.  
We let the predictions of the authors stand as originally made.

Occasionally, new authors have revised the predictions of older
authors.  We simply include both predictions.

All non-Standard Model predictions concern $x$, the amplitude
for mixing through virtual intermediate states (in units of
one-half the mean $\DZ$ decay rate).  Most non-Standard Models
contain numerous adjustable parameters; generally, we have tried
to take values of the adjustable parameters that the authors themselves
recommend.

\begin{figure}[htpb]
\begin{center}
\epsfig{figure=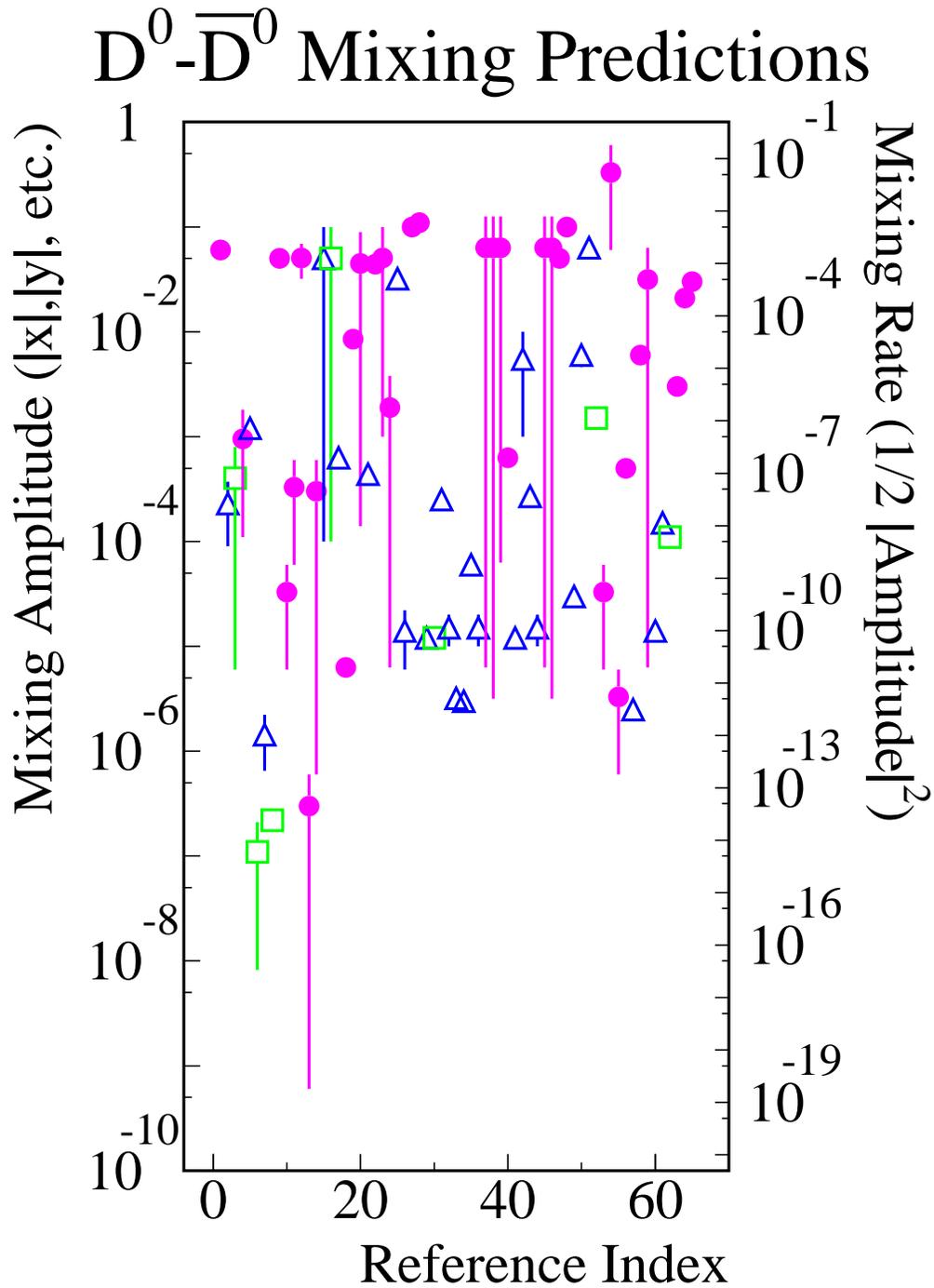,width=4.9in}
\end{center}
\caption[$\DZ\!-\!\DZB$ Mixing Predictions]
{$\DZ\!-\!\DZB$ mixing predictions; the vertical direction,
read off the left scale, is the mixing amplitude $x$, $y$,
or, if the appropriate strong phase is negligible, $x^\prime$,
$y^\prime$.  The right vertical scale is the equivalent
mixing \emph{rate},  which is either $(1/2)x^2$ or $(1/2)y^2$.
The horizontal is the Reference Index, which is a number assigned
to each prediction, and documented in Tables~\ref{tbl:ra}-\ref{tbl:rb}.
The open triangles (blue) are Standard Model predictions for $x$, the open
squares (green) are Standard Model predictions for $y$, and 
the solid circles (magenta) are non-Standard Models for $x$.}
\label{fig:pred}
\end{figure}

Our compilation of predictions are summarized in
Tables~\ref{tbl:ra}-\ref{tbl:rb} and Fig.~\ref{fig:pred}.

We assign a `Reference Index' to each prediction that we include,
and the first column of the Tables~\ref{tbl:ra}-\ref{tbl:rb} is that
Reference Index.  Roughly, the
reference index is assigned chronologically. 
The predictions are then plotted in Fig.~\ref{fig:pred}, with
the horizontal axis being this Reference Index.  

Some papers contain
multiple predictions.  Usually we assign each prediction a new
Reference Index.  However, a family of predictions is usually
assigned a range of predictions, and is plotted in Fig.~\ref{fig:pred}
as a central value and an error bar, where the error bar construes
the range, above a single Reference Index.

When mixing is measured with the decay of the $\DZ$ to a particular
hadronic final state $f_{\rm had}$, there is the possibility
of a $CP$-conserving, \emph{relative} strong phase $\delta$ between the 
amplitude for the decay $\DZ\!\to\!f_{\rm had}$, and that for
$\DZB\!\to\!f_{\rm had}$. 

In the useful case of $f_{\rm had}=K^+\pi^-$,
it has been argued\cite{brpa} that $\delta$ is small.
Although the
strong phase $\delta_I$, between the decay amplitudes 
for $\DZB\!\to\!K^+\pi^-$ to specific
isospin configurations of the $K^+\pi^-$ system,
is known to be large, the \emph{relative}
strong phase $\delta$ between the \emph{total} $\DZB\!\to\!K^+\pi^-$ and
$\DZ\!\to\!K^+\pi^-$ amplitudes is probably much smaller.

The presence of the relative
strong phase $\delta$ causes mixing studies  that use a specific hadronic
final state to be, in effect, sensitive to a \emph{rotated} set of mixing
amplitudes:
\begin{equation}
\begin{array}{rcl}
y^{\prime}&=&y\cos\delta - x\sin\delta\\
x^{\prime}&=&x\cos\delta + y\sin\delta
\end{array}.
\end{equation}
When $\delta$ is small, $y^{\prime}\approx y$, and $x^{\prime}\approx x$.

\begin{table}[p]
\caption[Reference Indices from 1 to 40]
{Reference Indices from 1 to 40.  The notation S stands
for `Standard Model' and the notation NS stands for
`non-Standard Model'.  The notation `$\pm$' does not indicate
a $1\,\sigma$ region, but an entire range of predictions, where
unknowable parameters govern the variation.}
\begin{center}
\begin{tabular}{c|cccc|c}
\textbf{Reference} & \textbf{Cita-} & 
 \textbf{Ampli-} & & &  \\ 
\textbf{Index} & \textbf{tion} & 
\textbf{tude} & \textbf{S/NS} & 
\textbf{Value} & \textbf{Comment} \\ 
\hline
\textbf{1} & \cite{vms} & $x$ & NS & $6\!\times\!10^{-2}$ &
\begin{minipage}[t]{2.2 in}
Family Symmetry
\end{minipage}\\ 

\textbf{2} & \cite{hyc} & $x$ & S & $(0.9\pm3.7)\!\times\!10^{-4}$ &
\begin{minipage}[t]{2.2 in}
Short Distance
\end{minipage}\\ 

\textbf{3} & \cite{hyc} & $y$ & S & $-(0.06\!-\!8.0)\times\!10^{-4}$ &
\begin{minipage}[t]{2.2 in}
Short Distance
\end{minipage}\\

\textbf{4} & \cite{hyc} & $x$ & NS & $(0.11\pm1.8)\!\times\!10^{-3}$ &
\begin{minipage}[t]{2.2 in}
Higgs Doublet
\end{minipage}\\

\textbf{5} & \cite{pst} & $x$ & S & $1.2\!\times\!10^{-3}$ &
\begin{minipage}[t]{2.2 in}
Short Distance
\end{minipage}\\

\textbf{6} & \cite{pst} & $y$ & S & $(0.082\!-\!2.1)\!\times\!10^{-7}$ &
\begin{minipage}[t]{2.2 in}
Short Distance
\end{minipage}\\

\textbf{7} & \cite{dk} & $x$ & S & $(1.44\pm0.79)\!\times\!10^{-6}$ &
\begin{minipage}[t]{2.2 in}
Short Distance
\end{minipage}\\

\textbf{8} & \cite{dk} & $y$ & S & $2.2\!\times\!10^{-7}$ &
\begin{minipage}[t]{2.2 in}
Short Distance
\end{minipage}\\

\textbf{9} & \cite{dat} & $x$ & NS & $5\!\times\!10^{-2}$ &
\begin{minipage}[t]{2.2 in}
Higgs Doublet
\end{minipage}\\

\textbf{10} & \cite{dat} & $x$ & NS & $(0.6\!-\!6.0)\!\times\!10^{-5}$ &
\begin{minipage}[t]{2.2 in}
L-R Symmetry
\end{minipage}\\

\textbf{11} & \cite{dat} & $x$ & NS & $(0.6\!-\!6.0)\!\times\!10^{-4}$ &
\begin{minipage}[t]{2.2 in}
Broken L-R Symmetry
\end{minipage}\\

\textbf{12} & \cite{dat} & $x$ & NS & $(5.05\pm1.85)\!\times\!10^{-2}$ &
\begin{minipage}[t]{2.2 in}
Kane-Thun Model
\end{minipage}\\

\textbf{13} & \cite{dat} & $x$ & NS & $(0.06\!-\!60)\!\times\!10^{-8}$ &
\begin{minipage}[t]{2.2 in}
SUSY
\end{minipage}\\

\textbf{14} & \cite{dat} & $x$ & NS & $(0.06\!-\!60)\!\times\!10^{-5}$ &
\begin{minipage}[t]{2.2 in}
SUSY - large CKM
\end{minipage}\\

\textbf{15} & \cite{wolf} & $x$ & S & $(0.01\!-\!10)\!\times\!10^{-2}$ &
\begin{minipage}[t]{2.2 in}
Long Distance
\end{minipage}\\

\textbf{16} & \cite{wolf} & $y$ & S & $(0.01\!-\!10)\!\times\!10^{-2}$ &
\begin{minipage}[t]{2.2 in}
Long Distance
\end{minipage}\\

\textbf{17} & \cite{dght} & $x$ & S & $6.3\!\times\!10^{-4}$ &
\begin{minipage}[t]{2.2 in}
Long Distance
\end{minipage}\\

\textbf{18} & \cite{eg} & $x$ & NS & $6.3\!\times\!10^{-6}$ &
\begin{minipage}[t]{2.2 in}
L-R Sym., $10\,$TeV Higgs
\end{minipage}\\

\textbf{19} & \cite{eg} & $x$ & NS & $8.5\!\times\!10^{-3}$ &
\begin{minipage}[t]{2.2 in}
FCNC, at $\KZ$ bound
\end{minipage}\\

\textbf{20} & \cite{mrr} & $x$ & NS & $(0.15\!-\!90)\!\times\!10^{-3}$ &
\begin{minipage}[t]{2.2 in}
Superstring-inspired $E_6$
\end{minipage}\\

\textbf{21} & \cite{cs} & $x$ & S & $4.4\!\times\!10^{-4}$ &
\begin{minipage}[t]{2.2 in}
Short Distance
\end{minipage}\\

\textbf{22} & \cite{cs} & $x$ & NS & $4.4\!\times\!10^{-2}$ &
\begin{minipage}[t]{2.2 in}
Higgs Doublets, $m_H\!=\!1\,$TeV
\end{minipage}\\

\textbf{23} & \cite{bhlp} & $x$ & NS & $(0.1\!-\!10)\!\times\!10^{-2}$ &
\begin{minipage}[t]{2.2 in}
Fourth Generation
\end{minipage}\\

\textbf{24} & \cite{joshi} & $x$ & NS & $(0.06\!-\!40)\!\times\!10^{-4}$ &
\begin{minipage}[t]{2.2 in}
FCNC, seesaw limit
\end{minipage}\\

\textbf{25} & \cite{cnp} & $x$ & S & $3.2\!\times\!10^{-2}$ &
\begin{minipage}[t]{2.2 in}
Long Distance
\end{minipage}\\

\textbf{26} & \cite{ors} & $x$ & S & $(1.4\pm0.8)\!\times\!10^{-5}$ &
\begin{minipage}[t]{2.2 in}
HQET
\end{minipage}\\

\textbf{27} & \cite{ns} & $x$ & NS & $\approx0.1$ &
\begin{minipage}[t]{2.2 in}
(s)/quark mass matrix align.
\end{minipage}\\

\textbf{28} & \cite{hw} & $x$ & NS & $0.11$ &
\begin{minipage}[t]{2.2 in}
Flavor Changing Scalar Int.
\end{minipage}\\

\textbf{29} & \cite{gburda} & $x$ & S & $1.2\times10^{-5}$ &
\begin{minipage}[t]{2.2 in}
Short Distance
\end{minipage}\\

\textbf{30} & \cite{gburda} & $y$ & S & $1.2\times10^{-5}$ &
\begin{minipage}[t]{2.2 in}
Short Distance
\end{minipage}\\

\textbf{31} & \cite{gburda} & $x$ & S & $2.5\times10^{-4}$ &
\begin{minipage}[t]{2.2 in}
Dispersive
\end{minipage}\\

\textbf{32} & \cite{gburda} & $x$ & S & $(1.5\pm0.5)\times10^{-5}$ &
\begin{minipage}[t]{2.2 in}
HQET
\end{minipage}\\

\textbf{33} & \cite{pkva} & $x$ & S & $3.2\times10^{-6}$ &
\begin{minipage}[t]{2.2 in}
Short Distance
\end{minipage}\\

\textbf{34} & \cite{hew} & $x$ & S & $3.0\times10^{-6}$ &
\begin{minipage}[t]{2.2 in}
Short Distance
\end{minipage}\\

\textbf{35} & \cite{hew} & $x$ & S & $6.0\times10^{-5}$ &
\begin{minipage}[t]{2.2 in}
Dispersive
\end{minipage}\\

\textbf{36} & \cite{hew} & $x$ & S & $(1.5\pm0.5)\times10^{-5}$ &
\begin{minipage}[t]{2.2 in}
HQET
\end{minipage}\\

\textbf{37} & \cite{hew} & $x$ & NS & $(0.006\!-\!120)\times10^{-3}$ &
\begin{minipage}[t]{2.2 in}
$4^{\rm th}$ Generation
\end{minipage}\\

\textbf{38} & \cite{hew} & $x$ & NS & $(0.004\!-\!120)\times10^{-3}$ &
\begin{minipage}[t]{2.2 in}
Higgs Doublet
\end{minipage}\\

\textbf{39} & \cite{hew} & $x$ & NS & $(0.06\!-\!120)\times10^{-3}$ &
\begin{minipage}[t]{2.2 in}
Flavor-Changing Higgs
\end{minipage}\\

\textbf{40} & \cite{bp} & $x$ & NS & $6.3\times10^{-4}$ &
\begin{minipage}[t]{2.2 in}
Isosinglet Quarks
\end{minipage}\\

\hline
\end{tabular}
\end{center}
\label{tbl:ra}
\vfill
\end{table}

\begin{table}[p]
\caption[Reference Indices from 41 to 65.]
{Reference Indices from 41 to 65.  The notation S stands
for `Standard Model' and the notation NS stands for
`non-Standard Model'.  The notation `$\pm$' does not indicate
a $1\,\sigma$ region, but an entire range of predictions, where
unknowable parameters govern the variation.}
\begin{center}
\begin{tabular}{c|cccc|c}
\textbf{Reference} & \textbf{Cita-} & 
 \textbf{Ampli-} & & &  \\ 
\textbf{Index} & \textbf{tion} & 
\textbf{tude} & \textbf{S/NS} & 
\textbf{Value} & \textbf{Comment} \\ 
\hline
\textbf{41} & \cite{gburdb} & $x$ & S & $5.8\!\times\!10^{-5}$ &
\begin{minipage}[t]{2.2 in}
Short Distance
\end{minipage} \\

\textbf{42} & \cite{gburdb} & $x$ & S & $(1\!-\!10)\!\times\!10^{-3}$ &
\begin{minipage}[t]{2.2 in}
Long Distance
\end{minipage} \\

\textbf{43} & \cite{gburdb} & $x$ & S & $2.7\!\times\!10^{-4}$ &
\begin{minipage}[t]{2.2 in}
Dispersive
\end{minipage}\\ 

\textbf{44} & \cite{gburdb} & $x$ & S & $(1.5\pm0.5)\times10^{-5}$ &
\begin{minipage}[t]{2.2 in}
HQET
\end{minipage}\\

\textbf{45} & \cite{gburdb} & $x$ & NS & $(0.06\!-\!120)\times10^{-3}$ &
\begin{minipage}[t]{2.2 in}
$4^{\rm th}$ Generation
\end{minipage}\\

\textbf{46} & \cite{gburdb} & $x$ & NS & $(0.04\!-\!120)\times10^{-3}$ &
\begin{minipage}[t]{2.2 in}
Higgs Doublet
\end{minipage}\\

\textbf{47} & \cite{gburdb} & $x$ & NS & $5\times10^{-2}$ &
\begin{minipage}[t]{2.2 in}
Tree Level FCNC
\end{minipage}\\

\textbf{48} & \cite{gburdb} & $x$ & NS & $0.1$ &
\begin{minipage}[t]{2.2 in}
SUSY
\end{minipage}\\

\textbf{49} & \cite{kaed} & $x$ & S & $3\times10^{-5}$ &
\begin{minipage}[t]{2.2 in}
Short Distance
\end{minipage}\\

\textbf{50} & \cite{kaed} & $x$ & S & $(6.0\pm1.4)\times10^{-3}$ &
\begin{minipage}[t]{2.2 in}
Broken SU(3), Octet
\end{minipage}\\

\textbf{51} & \cite{kaed} & $x$ & S & $6\times10^{-2}$ &
\begin{minipage}[t]{2.2 in}
Upper Limit
\end{minipage}\\

\textbf{52} & \cite{blp} & $y$ & S & $1.5\times10^{-3}$ &
\begin{minipage}[t]{2.2 in}
Phenomenological B. R.'s
\end{minipage}\\

\textbf{53} & \cite{ars} & $x$ & NS & $(0.6\!-\!6)\times10^{-5}$ &
\begin{minipage}[t]{2.2 in}
Higgs Doublet
\end{minipage}\\

\textbf{54} & \cite{ars} & $x$ & NS & $(0.6\!-\!6)\times10^{-1}$ &
\begin{minipage}[t]{2.2 in}
Higgs Doublet
\end{minipage}\\

\textbf{55} & \cite{ars} & $x$ & NS & $(0.6\!-\!6)\times10^{-6}$ &
\begin{minipage}[t]{2.2 in}
Higgs Doublet
\end{minipage}\\

\textbf{56} & \cite{ckln} & $x$ & NS & $5\times10^{-4}$ &
\begin{minipage}[t]{2.2 in}
SUSY
\end{minipage}\\

\textbf{57} & \cite{petr} & $x$ & S & $2.5\times10^{-6}$ &
\begin{minipage}[t]{2.2 in}
Dipenguin
\end{minipage}\\

\textbf{58} & \cite{tmat} & $x$ & NS & $6\times10^{-4}$ &
\begin{minipage}[t]{2.2 in}
Neutral Scalar Subquarks
\end{minipage}\\

\textbf{59} & \cite{ky} & $x$ & NS & $(0.06\!-\!600)\times10^{-4}$ &
\begin{minipage}[t]{2.2 in}
Singlet Quarks
\end{minipage}\\

\textbf{60} & \cite{abnp} & $x$ & S & $1.4\times10^{-5}$ &
\begin{minipage}[t]{2.2 in}
U(4)$_{\rm L}\times$U(4)$_{\rm R}$ Chiral
\end{minipage}\\

\textbf{61} & \cite{gp} & $x$ & S & $1.5\times10^{-4}$ &
\begin{minipage}[t]{2.2 in}
Long Dist. - Resonances
\end{minipage}\\

\textbf{62} & \cite{gp} & $y$ & S & $1.1\times10^{-4}$ &
\begin{minipage}[t]{2.2 in}
Long Dist. - Resonances
\end{minipage}\\

\textbf{63} & \cite{bcfpt} & $x$ & NS & $3\times10^{-3}$ &
\begin{minipage}[t]{2.2 in}
FCNC Dualized SM
\end{minipage}\\

\textbf{64} & \cite{kk} & $x$ & NS & $2.1\times10^{-2}$ &
\begin{minipage}[t]{2.2 in}
SUSY Broken by Flavor
\end{minipage}\\

\textbf{65} & \cite{bn} & $x$ & NS & $3.0\times10^{-2}$ &
\begin{minipage}[t]{2.2 in}
Flavor Changing from Higgs
\end{minipage}\\ \hline

\end{tabular}
\end{center}
\label{tbl:rb}
\vspace{3in}
\end{table}

\vfill

\begin{thebibliography}{99}
\bibitem{RPP98} C.~Caso, \emph{et. al.} European Physical Journal
\textbf{C 3} 1 (1998).

\bibitem{brpa}T.~E.~Browder and S.~Pakvasa, Phys. Lett. {\bf B 383},
475 (1996), \texttt{hep-ph/9508362}.

\bibitem{vms}G.~Volkov, V.~A.~Monich, and B.~V.~Struminski,
Yad. Fiz. {\bf 34}, 435 (1981).

\bibitem{hyc}H.-Y.~Cheng,
Phys. Rev. {\bf D 26}, 143 (1982).

\bibitem{pst}E.~A.~Paschos, B.~Stech, and U.~T\"urke,
Phys. Lett. {\bf B 128}, 240 (1983).

\bibitem{dk}A.~Datta and D.~Kumbhakar,
Z. Phys. {\bf C 27}, 515 (1985).

\bibitem{dat}A.~Datta,
Phys. Lett. {\bf B 154}, 287 (1985).

\bibitem{wolf}L.~Wolfenstein
Phys. Lett. {\bf B 164}, 170 (1985).

\bibitem{dght}J.~F.~Donoghue, E.~Golowich, B.~R.~Holstein,
and J.~Trampeti\'c,
Phys. Rev. {\bf D 33}, 179 (1986).

\bibitem{eg}G.~Ecker and W.~Grimus,
Z. Phys. {\bf C 30}, 293 (1986).

\bibitem{bgz}I.~I.~Bigi, G.~K\"opp, and P.~M.~Zerwas,
Phys. Lett. {\bf B 166}, 238 (1986).

\bibitem{mrr}
B.~Mukhopadhyaya, A.~Raychaudhuri, and A.~Ray-Mukhopadhyaya,
Phys. Lett. {\bf B 190}, 93 (1987).

\bibitem{cs}
T.~P.~Cheng and M.~Sher,
Phys. Rev. {\bf D 35}, 3484 (1987).

\bibitem{bhlp}
K.~S.~Babu, X.~G.~He, X.~Li, and S.~Pakvasa,
Phys. Lett. {\bf B 205}, 540 (1988).

\bibitem{joshi}
A.~S.~Joshipura,
Phys. Rev. {\bf D 39}, 878 (1989).

\bibitem{cnp}
P.~Colangelo, G.~Nardulli, and N.~Paver,
Phys. Lett. {\bf B 224}, 71 (1990).

\bibitem{ors}
T.~Ohl, G.~Ricciardi, and E.~Simmons,
Nucl. Phys. {\bf B 403}, 71 (1993);
\texttt{hep-ph/9301212}.

\bibitem{ns}
Y.~Nir and N.~Seiberg,
Phys. Lett. {\bf B 309}, 337 (1993);
\texttt{hep-ph/9304407}.

\bibitem{hw}
L.~Hall and S.~Weinberg,
Phys. Rev. {\bf D 48}, 979 (1993);
\texttt{hep-ph/9303241}.

\bibitem{gburda}G.~Burdman, in \emph{The Future of High-Sensitivity
Charm Experiments},
edited by Daniel M. Kaplan and Simon Kwan, (Fermilab, Batavia, Ill., 1994) pp. 75-84;
\texttt{hep-ph/9407378}.

\bibitem{pkva}
S.~Pakvasa,
Chin. J. Phys. {\bf 32}, 1163 (1994);
\texttt{hep-ph/9408270}.

\bibitem{hew}J.~L.~Hewett, in \emph{The Albuquerque Meeting},
edited by Sally Seidel, (World Scientific, River Edge, N.J., 1995) pp. 951-955;
\texttt{hep-ph/9409379}.

\bibitem{bp}
G.~C.~Branco, P.~A.~Parada,
Phys. Rev. {\bf D 52}, 4217 (1995);
\texttt{hep-ph/9501347}.

\bibitem{gburdb}G.~Burdman, in \emph{Workshop on the Tau/Charm Factory},
edited by Jose Repond, (AIP Press, Woodbury, N.Y., 1995) pp. 409-424;
\texttt{hep-ph/9508349}.

\bibitem{kaed}
T.~A.~Kaeding,
Phys. Lett. {\bf B 357}, 151 (1995);
\texttt{hep-ph/9505393}.

\bibitem{blp}
F.~Buccella, M.~Lusignoli, and A.~Pugliese,
Phys. Lett. {\bf B 379}, 249 (1996);
\texttt{hep-ph/9601343}.

\bibitem{ars}
D.~Atwood, L.~Reina, and A.~Soni,
Phys. Rev. {\bf D 55}, 3156 (1997);
\texttt{hep-ph/9609279}.

\bibitem{ckln}
A.~G.~Cohen, D.~B.~Kaplan, F.~Lepeintre, and A.~E.~Nelson,
Phys. Rev. Lett. {\bf 78}, 2300 (1997);
\texttt{hep-ph/9610252}.

\bibitem{petr}
A.~A.~Petrov,
Phys. Rev. {\bf D 56}, 1685 (1997);
\texttt{hep-ph/9703335}.

\bibitem{tmat}
T.~Matsushima,
\texttt{hep-ph/9704316}.

\bibitem{ky}
I.~Kakebe and K.~Yamamoto,
Phys. Lett. {\bf B 416}, 184 (1998);
\texttt{hep-ph/9705203}.

\bibitem{abnp}
G.~Amoros, F.~J.~Botella, S.~Noguera, and J.~Portoles,
Phys. Lett. {\bf B 422}, 265 (1998);
\texttt{hep-ph/9707293}.

\bibitem{gp}
E.~Golowich and A.~A.~Petrov
Phys. Lett. {\bf B 427}, 172 (1998);
\texttt{hep-ph/9802291}.

\bibitem{bcfpt}
J.~Bordes, H.-M.~Chan, J~Faridani, J~Pfaudler, and S.-T.~Tsou,
Phys. Rev. {\bf D 60}, 013005 (1999);
\texttt{hep-ph/9807277}.

\bibitem{kk}
D.~E.~Kaplan and G.~D.~Kribs,
\texttt{hep-ph/9906341}.

\bibitem{bn}
K.~S.~Babu and S.~Nandi, 
\texttt{hep-ph/9907213}.

\end{thebibliography}
\end{document}